\documentclass[12pt]{article}
\usepackage{epsfig} \usepackage{amssymb} \usepackage{amsfonts}
\def\eV{{\rm e\kern-0.12em V}} \def\GeV{{\rm G}\eV} 
\def\MSbar{\relax\ifmmode\overline{\rm MS}\else{$\overline{\rm MS}${ }}\fi}
\def\msbar{\relax\ifmmode\overline{\rm MS}\else{$\overline{\rm MS}${ }}\fi}
\def\alphan{\relax\ifmmode{\alpha_{\rm an}}\else{$\alpha_{\rm an}${ }}\fi}
\def\alphas{\relax\ifmmode{\alpha_{\rm s}}\else{$\alpha_{\rm s}${ }}\fi}
\def\asmz{\relax\ifmmode\bar \alpha_s(M_Z^2)\else{$\bar \alpha_s(M_Z^2)${ }}\fi}
\def\albars{\relax\ifmmode{\bar{\alpha}_s}\else{$\bar{\alpha}_s${ }}\fi}
\def\tildal{\relax\ifmmode{\tilde{\alpha}}\else{$\tilde{\alpha}${ }}\fi}
\def\tildals{\relax\ifmmode{\tilde{\alpha}(s)}\else{$\tilde{\alpha}(s)${ }}\fi}
\def\asQ{\relax\ifmmode\bar{\alpha}_s(Q^2)\else{$\bar{\alpha}_s(Q^2)${ }}\fi}
\def\agoth{\relax\ifmmode{\mathfrak A}\else{${\mathfrak A}${ }}\fi}
\def\pisq{\relax\ifmmode{\pi^2}\else{${\pi^2}${ }}\fi}
\def\agothk{\relax\ifmmode{\mathfrak A}_k\else{${\mathfrak A}_k${ }}\fi}
\def\acal{\relax\ifmmode{\cal A}\else{${\cal A}${ }}\fi}
\def\acalk{\relax\ifmmode{\cal A}_k\else{${\cal A}_k${ }}\fi}
\newcommand{\beq}{\begin{equation}} \newcommand{\eeq}{\end{equation}}
\newcommand{\beglab}{\begin{equation}\label}
\begin{document}   

\begin{center}
{\large\sf Analytic Perturbation Theory and New Analysis of some
QCD observables \\}
\bigskip

D.V. Shirkov \
\medskip

{\it Bogoliubov Laboratory, JINR, 141980 Dubna, Russia\\
e-address: shirkovd@thsun1.jinr.ru}
\end{center} \smallskip

\abstract{Here, we report briefly two topics: \par
 1) The latest version of ``Analytic Perturbation Theory" (APT)
devised recently for the QCD observables both in the Euclidean and
Minkowskian regions. \par
 2) Results of the APT--based calculation for some physical
 processes.}  \smallskip
\section{The APT --- a closed theoretical scheme}
 The new APT version \cite{tow00,apt01} mutually relates two ghost--free
formulations of modified perturbation expansions for observables.
\par
 The first one, initiated about two decades ago \cite{rad82,kras82},
changes the standard power expansion in the time-like region
$$
R_{\rm pt}(s)=1+r_{\rm pt}(s)\, ;\quad  r_{\rm pt}(s)=\sum_{k\geq
1}r_k\, \alphas^k(s;f)\,$$
into a nonpower one $\/\: r_{\rm pt}(s) \to  r_{\pi}(s)
= \sum_{k\geq 1}d_k\, {\mathfrak A}_k(s,f) \,.$

Here, $\alphas(s)$ is a common, e.g., 3--loop QCD invariant/running
coupling (usually in the \msbar scheme -- see  eq.(9.5a) in Ref.
\cite{pdg00}); and $\agothk$, some integral images of the \alphas powers:
\beq  \mathfrak{A}_k(s)={\bf R} \left[\albars^k(Q^2)\right]\,; \quad
R(s)=\frac{i}{2\pi}\,\int^{s+i\varepsilon}_{s-i\varepsilon}\frac{d
z}{z}\, D(-z)\equiv{\bf R}\left[D_{\rm }(Q^2)\right]\,.\eeq

 The operation ${\bf R}$ is a reverse ${\bf R} = [{\bf D}]^{-1}$
to the one defined by the  ``Adler relation"
\beq\label{d-trans}
R(s) \to D(Q^2)=Q^2\int^{\infty}_0 \frac{d s}{(s+Q^2)^2}\,R(s)\,
                \equiv {\cal\bf D} \left\{ R(s)\right\}\, \eeq
\noindent and transforming a real function $\,R(s)\,$ of a positive
(time--like) argument into a real function $\,D(Q^2)\,$ of a positive
(space--like) argument.\par

By operation ${\cal\bf R}\,,$ one can define \cite{rad82,kras82,js95}
the RG--invariant effective coupling
$\tildal(s)={\bf R}\left[\albars \right] \,$ in the time--like region. 
   A few simple examples are in order :     \smallskip

---  For the one--loop case  with $\albars^{(1)}=\left[\beta_0
\ln(Q^2/\Lambda^2)\right]^{-1}\,$ one has \cite{schr2,rad82,js95}
$$ {\bf R}\left[\albars^{(1)}\right]\,=
\agoth_1^{(1)}(s)=\frac{1}{\beta_0}
\left[\frac{1}{2}-\frac{1}{\pi}\arctan\frac{L}{\pi}\right]_{L>0}=\frac{1}
{\beta_0\pi}\arctan\frac{\pi}{L}\, ; \quad  L=\ln \frac{s}{\Lambda^2}
\,.\,\,\eqno{(3a)}$$

--- Square and cube of $\albars^{(1)}\,$ transform into simple expressions
\cite{rad82,kras82}
$$\agoth_2^{(1)}(s)\equiv {\bf R}\left[\left(\albars^{(1)} \right)^2\right]
=\frac{1}{\beta_0^2\left[L^2+\pi^2\right]}\,\quad \mbox{and} \quad
\agoth_3^{(1)}(s)=\frac{L}{\beta_0^3\left[L^2+\pi^2\right]^2}\,, \,\,
\eqno{(3b)}$$
(related by differential operation
$k\beta_0\agoth_{k+1}^{(1)}= -(d/dL)\agothk^{(1)}$)
which {\it are not powers} of  $\agoth^{(1)}_1\,.$ \par
\addtocounter{equation}{1}

By applying  ${\bf D}$ to  $\agoth_k(s)$ one can  ``try to return" to the
Euclidean domain. However, instead of \alphas powers, we arrive at some other
functions $\acalk(Q^2)={\bf D}\left[\agothk \right]\,,$  analytic in the cut
$Q^2$-plane and free of ghost singularities.   At the one--loop case

\beq\label{4} \beta_0 \acal^{(1)}_1(Q^2) = \frac{1}{\ln (Q^2/\Lambda^2)} -
\frac{\Lambda^2}{Q^2-\Lambda^2}\,,\quad   \beta_0^2
\acal^{(1)}_2(Q^2) = \frac{1}{\ln^2 (Q^2/\Lambda^2)} +
\frac{Q^2\Lambda^2}{(Q^2-\Lambda^2)^2}\,,\,\,\, \dots \,.\eeq

 These expressions have been first obtained by other means \cite{rapid96,prl97}
at mid--90s. The first function $\acal_1= \alphan(Q^2)\,,$ an {\it invariant
Euclidean coupling}, should now be treated as a
counterpart of the {\it invariant  Minkowskian coupling} \cite{js95}
$\tildal(s)=\agoth_1(s)\,.$ Both $\alphan$ and $\tildal$ are real
monotonically decreasing functions with the same maximum value
$\alphan(0)=\tildal(0)= 1/\beta_0({f=3})\simeq1.4\,$ in the IR limit.
 All higher functions vanish, $\acalk(0)=\agothk(0) =0$ in
this limit. For $k\geq 2\,,$ they oscillate in the IR region.  \par
 The same properties remain valid for a higher--loop case. Explicit
expressions for \acalk and \agothk at the two--loop case can be written
(see, Ref. \cite{mag00}) in terms of a special Lambert function. They are
presented in Figs 1a and 1b. \medskip

 \begin{figure}[th]
 \unitlength=1mm
   \begin{picture}(0,61)                                   %
   \put(0,1){
   \epsfig{file=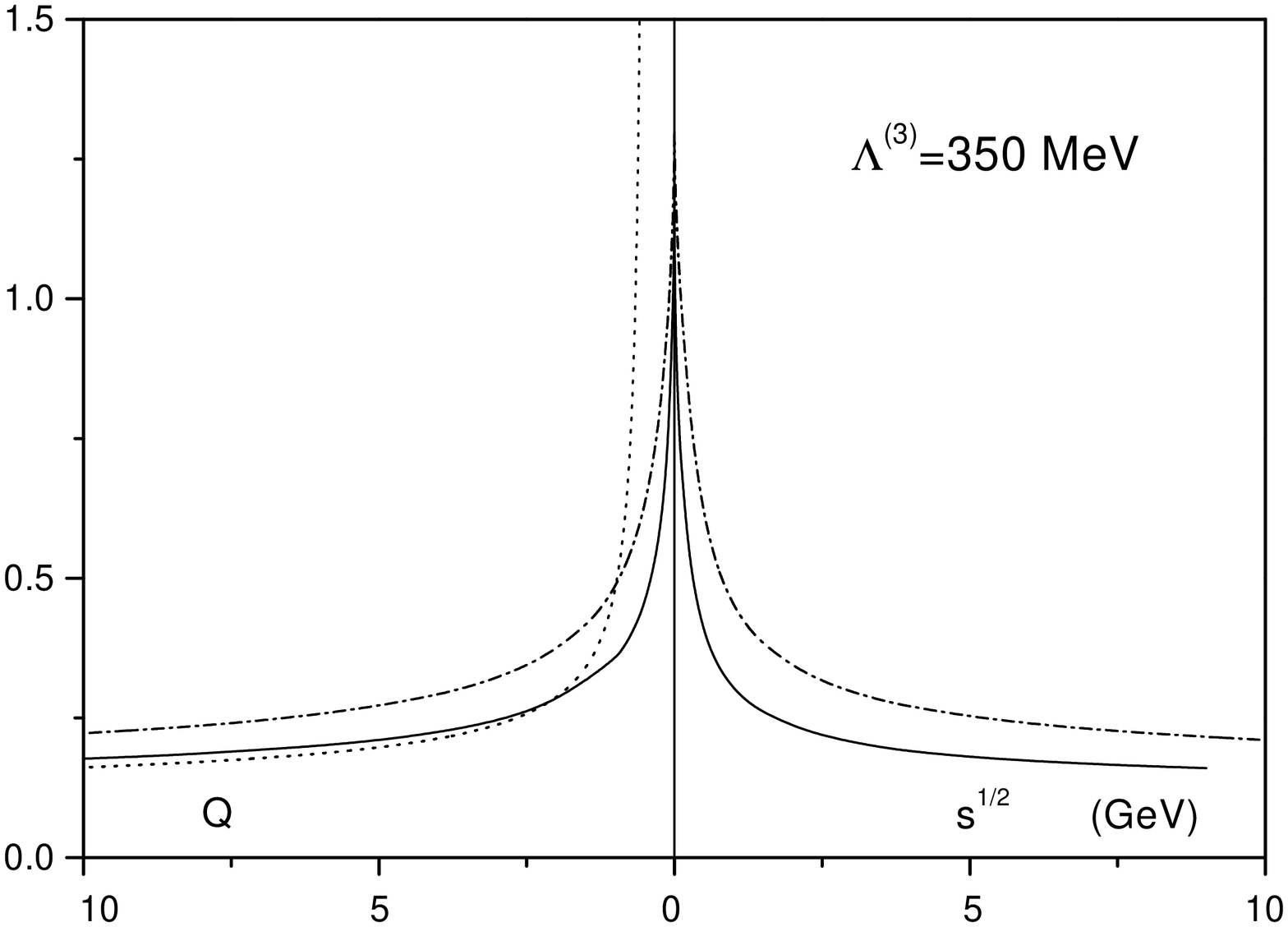,width=8.5cm}
   }
   \put(80,1){%
   \epsfig{file=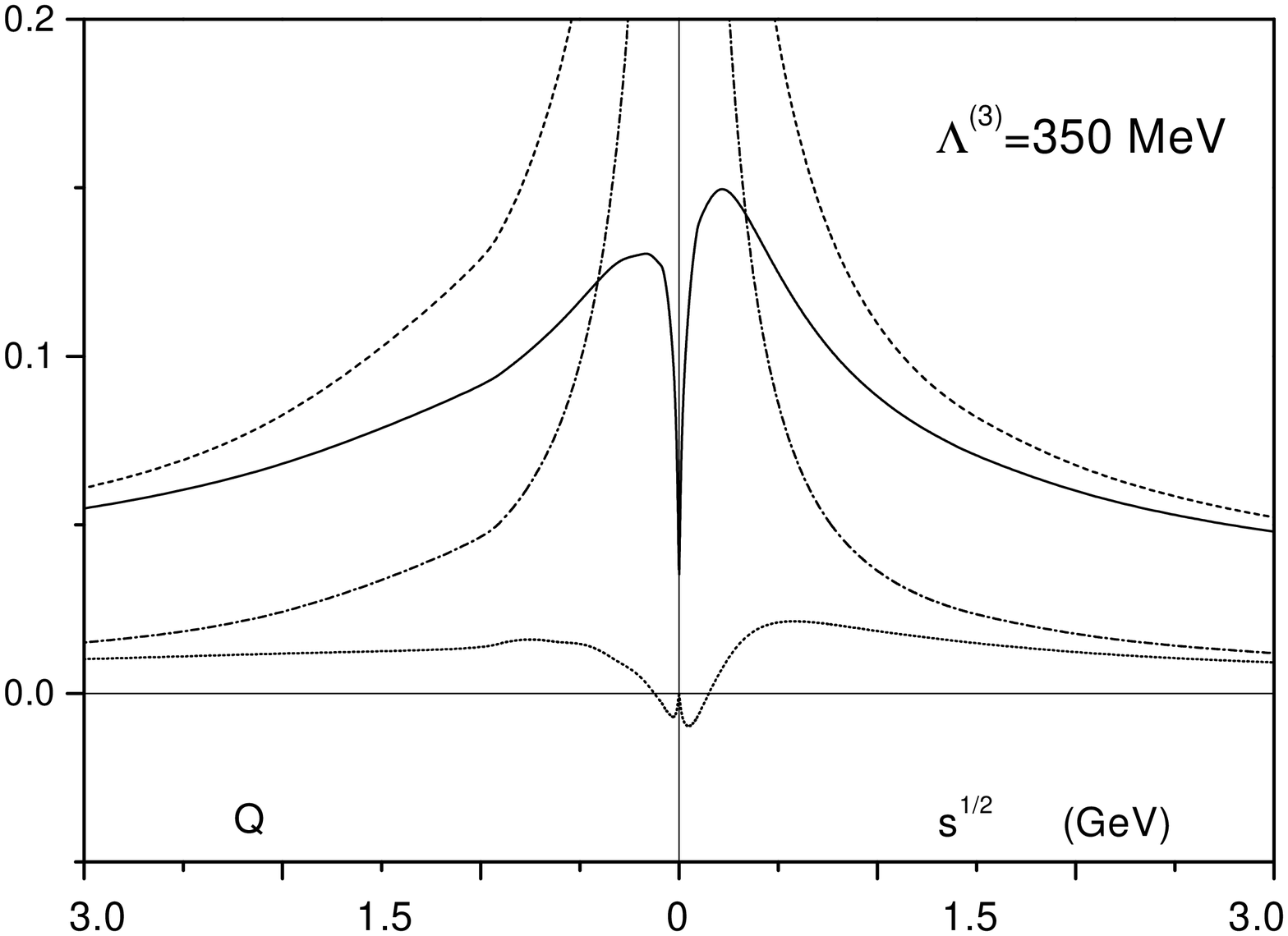,width=8.5cm}%
   }       %
 \put(14,53){\bf a}             %
 \put(94,53){\bf b}             %
  \put(43.8,53){$\bullet$}
  \put(27,53){\small $\alphan(0)=$}
  \put(48,53){\small \tildal(0)}
   \put(24,43){\small $\bar\alpha_s^{(2)}(Q^2;3)$}
   \put(19,23){\small  $\alpha_{\rm{an}}^{(1)}(Q^2;3)$}
   \put(33,14){\small $\alpha_{\rm an}^{(2,3)}$}
   \put(51,23){\small $\tilde{\alpha}^{(1)}(s;3)$}
   \put(46,13){\small $\tildal^{(2,3)}$}
    \put(110,50){\small $\alpha_{\rm an}^2$}
    \put(112,39){\small\bf ${\cal A}_2$}
    \put(101,26){\small $\alpha_{\rm an}^3$}
    \put(116,24){\small\bf ${\cal A}_3$}
    \put(138,39){\small $\tilde{\alpha}^2$}
    \put(133,33){\small ${\mathfrak A}_2$}
    \put(145,24){\small  $\tilde{\alpha}^3$}
    \put(127,25){\small ${\mathfrak A}_3$}
  \end{picture}
\centerline{    \parbox{15.5cm}{
\caption{\sl\footnotesize {\bf a} Space-like and time-like global
analytic couplings in a few GeV domain; {\bf b} ``Distorted
mirror symmetry" for global expansion functions. All the curves in
{\bf 1b} are given for the 2--loop case.} \label{fig1}
}}
 \end{figure}

Here, in Fig.1a, by the dotted line we  give a usual two-loop effective
QCD coupling \asQ with a pole at $Q^2=\Lambda^2\,.$ On the other hand, the
dash--dotted curves represent the one-loop APT approximations (3a) and
(\ref{4}). The solid APT curves are based on the exact two-loop solutions
of RG equations and approximate three--loop solutions in the \msbar scheme.
Their remarkable coincidence (within the 1--2 per cent) demonstrates
reduced sensitivity of the APT with respect to higher--loops effects in
the whole Euclidean and Minkowskian regions from IR to UV limits.
Fig.1b shows higher functions calculated at the two--loop case. \par

  Remarkably enough, the mechanism of liberation of unphysical
singularities is quite different. While in the space-like domain
it involves nonperturbative, power in $Q^2$, structures, in the time-like
region, it is based only upon resummation of the ``$\pi^2$ terms".
Figuratively, (non-perturbative~!) {\it analyticization} \cite{tmp00} in
the $Q^2$--channel can be treated as a quantitatively distorted reflection
(under $Q^2\to s=- Q^2$) of (perfectly perturbative) $\pi^2$--resummation
in the $s$--channel. This effect of ``distorting mirror" first discussed in
\cite{mo98} is clearly seen in figures. \smallskip

 In a real case, the procedure of the threshold matching is in use. E.g.,
in the \msbar scheme with $\albars(Q^2=M^2_f; f-1) =\albars(Q^2=M^2_f; f)$
it defines a ``global" function
$$\albars(Q^2)  =\albars(Q^2;f)\,  \quad \mbox{at} \,\quad  M^2_{f-1}
\leq Q^2\leq  M^2_f\,,$$
continuous in the space-like region of positive $\,Q^2\,$ values with
discontinuity of derivatives at matching
points. To this there corresponds a discontinuous spectral density
\begin{equation}\label{discont}
 \rho_k(\sigma)=\rho_k(\sigma; 3) +
\sum_{f\geq 4}^{}\theta(\sigma-M_f^2) \left\{\rho_k(\sigma; f)-\rho_k(\sigma;
f-1) \right\} \end{equation} with $~\rho_k(\sigma; f)=\Im\,\albars^k (-\sigma, f)$ which
yields the smooth global Euclidean and spline--continuous global Minkowskian
expansion functions
\begin{equation} \label{globalAQ}
{\cal A}_k(Q^2) =\frac{1}{\pi}\int\limits_{0}^{\infty} \frac{d\sigma} {\sigma+x}
\:\rho_k(\sigma)\,; \quad \mathfrak{A}_k(s)=\int\limits^{\infty}_{s}\frac{d
\sigma}{\sigma} \rho_k(\sigma)\,. \end{equation}


\section{The APT applications.}
To illustrate a qualitative difference between our global APT scheme and
common practice of data analysis, we  consider a few examples. \par

 In the usual treatment --- see, e.g., Ref. \cite{pdg00} --- the (QCD
perturbative part of)  Minkowskian observable, like $e^+e^-$ annihilation or
$Z_0$ decay, is presented in the form
\begin{equation}\label{Rtrad} \frac{R(s)}{R_0} =1+
r(s)\,; \quad r_{PT}(s)=\frac{\albars(s)}{\pi}+r_2\,
\albars^2(s)+ r_3\,\albars^3(s)\,. \end{equation}

 Here, coefficients $\, r_1=1/\pi\,,~\,r_2\, $ and $~r_3\,$ usually are not
diminishing.  A  rather big negative $r_3$ value comes mainly from
the $\,-r_1\pi^2\beta^2_0/3\,\,$ term. In the APT, we have instead
 \begin{equation}\label{rapt} r_{APT}(s)=d_1\tildal(s)+d_2\,{\mathfrak
A}_2(s) +d_3\:{\mathfrak A}_3(s)\:\end{equation}
with reasonably decreasing coefficients $\,d_1=r_1,~d_2=r_2\, $ and  $~d_3=r_3
+r_1\pi^2\beta^2_0/3\,,$ the mentioned $\pi^2$ term of $r_3$ being
``swallowed" by $\,\tildal(s)\,.$ \medskip

In the Euclidean channel, instead of a power expansion similar to
(\ref{Rtrad}), we typically have
 \begin{equation}\label{dapt}
d_{APT}(Q^2)=d_1\alpha_{\rm an}(Q^2) +d_2\,{\cal A}_2(Q^2)+
 d_3\,{\cal A}_3(Q^2)\:. \end{equation}
 Here, the modification is related to nonperturbative structures  like
in (\ref{4}).

\begin{center}
 {\sf Table 1 : {\small Relative contributions (in \%)
of 1-- , 2-- and 3--loop terms to observables}}
\label{tab1}\smallskip

\begin{tabular}{|l||c|c|c||r|r|r|}  \hline
\multicolumn{1}{|l||}{\slshape \phantom{aa}Process} &
\multicolumn{3}{c||}{\slshape PT}
 & \multicolumn{3}{r|}{\slshape APT\phantom{aaaa}}  \\    \hline\hline
GLS sum rule         & 65 & 24 & 11  &  75 & 21 & \bf{4} \\
\hline
Bjorken. s.r.         & 55 & 26 & 19  &  80 & 19 & \bf{1}  \\
\hline\hline Incl. $\tau$-decay     & 55 & 29 & 16  &  88 & 11 &
\bf{1}\\ \hline
$e^+e^- \to$ hadr.  & 96 &  8 & -4  &  92 &  7 & \bf{.5} \\
  (at 10 GeV) & & & & & & \\ \hline
$Z_o \to$ hadr.  & 98.6 & 3.7 & -2.3 & 96.9 & 3.5 & -\bf{.4} \\
\hline \end{tabular} \end{center} \medskip

In Table 1, we give values of the relative contribution of the first,
second, and third terms of the r.h.s. in (\ref{Rtrad}),(\ref{rapt}) and
(\ref{dapt}) for Gross--Llywellin-Smith \cite{mss99} and Bjorken \cite{mss98}
sum rules, $\tau$ -- decay in the vector channel \cite{mss01} as well as
for $e^+e^-$ and $Z_0$ inclusive cross-sections. As it follows from this
Table, in the APT case, the three--loop (last) term is very small, and
being compared with data errors, numerically unessential. This means that,
in practice,

\centerline{\sf\normalsize one can use the APT expansions (\ref{rapt}) and
(\ref{dapt}) without the last term.}  \smallskip

 Using this conclusion as a hint, we reanalyzed data in the five--flavor
region. Results are presented in Table 2.  \par  \medskip

\begin{center}
\begin{minipage}[t]{155mm}
\begin{center}
{\sf Table 2 :  {The APT revised
part ($f=5$) of Bethke's \cite{beth00} Table 6 }} \\
\scriptsize{ Figures in brackets  give the  difference
 in the last digit between APT and common  \asmz values.}\smallskip

 \normalsize
\begin{tabular}{|c|c|c||c|c||c|c|}  \hline 
&$\sqrt{s}$&loops&\albars(s)&\asmz&\albars(s)&\asmz\\
Process&\GeV& No& ref.[2] &ref.[2]& APT &APT  \\ \hline \hline
$\Upsilon$-decay \cite{pdg00} &9.5&2 &.170&.114 &.182&.120 (+6) \\
$e^+e^-[\sigma_{had}]$&10.5&{\bf 3}&.200&.130 &.198&.129({\bf -1}) \\ 
$e^+e^-[j\, \&\, sh]$   &22.0&2 &.161&.124 &.166&.127(+3)  \\
$e^+e^-[j\, \&\, sh]$   &35.0&2 &.145&.123 & .149&.126(+3)  \\ 
$e^+e^-[\sigma_{had}]$&42.4&{\bf 3} &.144&.126 &.145&.127(+{\bf1}) \\ 
$e^+e^-[j\, \& sh]$     &44.0&2 &.139&.123 &.142&.126(+3) \\ 
$e^+e^-[j\, \& sh]$     &58  &2 &.132&.123 &.135&.125(+2)  \\ 
{\bf $Z_0\to$ had.}     &91.2&{\bf 3} &.124&.124 &.124&.124 ({\bf 0})  \\
$e^+e^-[j\,\&\,sh]$     &91.2&2 &.121&.121&.123&.123(+2) \\
 -"-                         & ....  & 2 & ...  & ...   & ...   & ... (+2)\\
$e^+e^-[j\,\&\,sh]$     &189 &2 &.110&.123& .112&.125(+2) \\
\hline   \end{tabular}
\end{center}
\end{minipage}
Averaged $<\asmz>_{f=5}$ values = \/$0.121$
\hspace{16mm} $0.124\,.$ \\
\end{center}
\vspace{2mm} \medskip

Addressing the reader interested in a more detail to our recent paper
\cite{pi2-00}, we shortly comment that transition PT $\to$ APT from the
standard algorithm to our new one for the NLO case, as a rule, enlarges
extracted \asmz values by 0.002 (or more) which results in the averaged
\asmz value equal to 0.124.  \par
  At the same time it improves the correlation of events in the $f=5$
region.  More specifically, it changes the $ <\chi^2>_{f=5}$ value from
0.197 to 0.144.

\section{Summary.}

 1. First, we have outlined  a  recently devised self-consistent
scheme for analyzing data both in the space-like and time-like
regions. Within this APT scheme, perturbative expressions for an
observable involve expansions over the sets $\,\left\{{\cal
A}_k(Q^2)\right\}\,$ and $\,\left\{\agothk(s)\right\}\,,$ that are
nonpower series, free of unphysical singularities, with usual
numerical coefficients $\,d_k\,$ obtained by calculation of the
relevant Feynman diagrams. \par

 2.  Numerically, the APT calculations reveal reduced sensitivity to the
NNLO effects, as it has been demonstrated in Table 1. \par
    Table 2 summarizes our attempt to ``improve" some particular data for
\asmz values extracted from experiments in  the Minkowskian  five-flavour
region. These results look encouraging. In particular, they yield the new
value
$$ <\asmz >_{f=5}= 0.124\,, $$
quite different from the widely accepted  ``world average"  0.118  and even
from the usual average ($= 0.121$) over the  five--flavour region.
\smallskip

3. This result, being taken as granted, rises a physical question on mutual
consistency of current data on the QCD--invariant coupling behavior in the
``medium ($f=3,4$)" and  ``high $(f=5,6)$"  regions.  \par
 Answer to this question could be obtained by a further  revised (within
the APT technique) calculation.
\smallskip

\section*{Acknowledgments}

This work was partially supported by grants of the Russian Foundation for
Basic Research (RFBR projects Nos 99-01-00091 and 00-15-96691),
by INTAS grant No 96-0842 and by INTAS-CERN grant No 2000-377.

\end{document}